\documentclass[12pt,a4paper]{article}
\usepackage{amssymb,amsmath,amscd,epsfig}
\title{Analysis of the SN1987A two-stage explosion hypothesis with account for the MSW neutrino flavour conversion.}
\author{
 Oleg Lychkovskiy 
\hspace*{2mm}$^{\rm a,b}$
\\ ${\rm ^a}$ {\small\it Institute for Theoretical and Experimental Physics}\\
{\small\it 117218, B.Cheremushkinskaya 25,
Moscow, Russia}\\
${\rm ^b}$ {\small\it Moscow Institute of Physics and Technology
}\\{\small\it 141700, 9, Institutskii per., Dolgoprudny, Moscow
Region, Russia}}

\date{}

\begin{document}

\newcommand{\nue}{\nu_e}
\newcommand{\nul}{\nu_l}
\newcommand{\antinue}{\bar{\nu}_e}
\newcommand{\antinul}{\bar{\nu}_l}
\newcommand{\nux}{\nu_x}
\newcommand{\antinux}{\bar{\nu}_x}
\newcommand{\ar}{\rightarrow}
\newcommand{\C}{{\rm C}}
\newcommand{\Hy}{{\rm H}}
\newcommand{\F}{{\rm F}}
\newcommand{\Co}{{\rm Co}}
\newcommand{\N}{{\rm N}}
\newcommand{\Ox}{{\rm O}}
\newcommand{\Fe}{{\rm Fe }}
\newcommand{\be}{\begin{equation}}
\newcommand{\ee}{\end{equation}}
\newcommand{\LN}{{\bf LN}~}
\newcommand{\LI}{{\bf LI}~}
\newcommand{\SA}{{\bf SA}~}

\maketitle

\begin{abstract}
Detection of 5 events by the Liquid Scintillation Detector (LSD)
on February, 23, 1987 was interpreted in the literature as the detection of neutrinos from the first stage of the two-stage
supernova collapse. 
We pose rigid constraints on the properties of the first stage of the collapse, taking into account neutrino flavour conversion due to the MSW-effect and general properties of supernova neutrino emission. The constraints depend on the unknown neutrino mass hierarchy and mixing angle $\theta_{13}.$
\end{abstract}


{Key words: {\it supernova neutrino; neutrino, mass spectrum; neutrino, mixing angle; \\SN1987A; neutrino; supernova.}}\\
\newline

 \section{Introduction}

SN1987A was the only supernova to date which produced a measured neutrino signal. Four experiments reported
the detection of neutrinos: LSD (Liquid Scintillator Detector, \cite{LSD1},\cite{LSD2}), KII (Kamiokande II, \cite{KII}), IMB (Irvine-Michigan-Brookheaven,\cite{IMB}) and BST (Baksan Scintillator Teleskope, \cite{Baksan}). While LSD registered neutrino burst at 2:52UT, February 23, the other three experiments -- at 7:35UT February 23 (UT stands for Unitary Time). Each experiment reported only {\it one} burst: LSD observed no statistically significant counterparts for the  KII, IMB and BST neutrino signals and vice versa. This puzzling discrepancy could be, in principle, explained by a two-stage supernova collapse hypothesis, as was stated in \cite{Hillebrandt}-
\cite{Imshennik}.
 Various two-stage supernova collapse models, proposed in this papers, implied the transition from a protoneutron star to a black hole and/or the formation and evolution of a close binary system inside the exploding star. 
Properties of the LSD signal and non-observation  of its counterpart in other detectors allow to pose rigid constraints
on the properties of the first stage of two-stage collapse scenarios, 
MSW effect~\cite{W,MS} in the matter of the star being of crucial importance.
For example, it was shown~\cite{Smirnov and Lunardini} that accounting for the neutrino flavour conversion 
can spoil the reported~\cite{Imshennik} concordance of the rotating collapsar model with the data. 
The analysis independent from the particular collapse model was made in \cite{Lychkovskiy}. In the present paper we extend the analysis of \cite{Lychkovskiy}: we conduct a more elaborate statistical study of the data and account for supernova shock wave effect, which may influence 
neutrino flavour conversion \cite{Schirato and Fuller}.  The obtained constraints on the first stage of the two-stage collapse depend on the unknown neutrino mass hierarchy and mixing angle 
$\theta_{13}.$ 


The structure of the paper is the following. In Section 2 we describe neutrino signals in the detectors involved. In Section 3 general properties of the supernova neutrino fluxes and flavour transformations are briefly reviewed. In Section 4 the analysis procedure and results are presented. In Section 5 the conclusions are summarised.

\section{SN1987A neutrino signal detection}
\subsection{General detector characteristics and SN1987A neutrino signals}

Detector characteristics and numbers of registered events at 2:52 UT and 7:35 are given in Table 1.
LSD reported their event cluster at 2:52 to be a burst, while
KII and BST regarded their few events at 2:52 as "background fluctuations". On the other hand,
KII, IMB and BST reported only event clusters at 7:35 to be "bursts". 
The duration of all bursts was about 10 seconds.


\begin{table}[t]
\caption{
Type, material and mass (in tons) of the detectors, numbers of events ($N_{\rm ev}$) at 2:52 and 7:35 
(according to the cited references). 
Only  events with   measured energy exceeding the energy thresholds 
of corresponding detectors were taken  into account.
\label{table 1}
			}
\vspace{0.2cm}
\begin{center}
 \begin{tabular}{|l|c|c|c|c|}
\hline
 & LSD & KII & IMB & Baksan\\
\hline
Type & scintillator & cherenkov & cherenkov & scintillator\\
\hline
Material  & $\C_n\Hy_{2n}~~~~90 {\rm t}$ & $\Hy_2\Ox~~~~2140 {\rm t}$ & $\Hy_2\Ox~~~~5000 {\rm t}$ & $\C_n\Hy_{2n}~~~~200 {\rm t}$\\
 & $\Fe~~~~~~~~200 {\rm t}$ & & &  \\
\hline
$N_{\rm ev}$ at 2:52 & 5~\cite{LSD1,LSD2} & 2~\cite{KII,DeRujula} & 0~\cite{IMB} & 1~\cite{Baksan}\\
\hline
$N_{\rm ev}$ at 7:35 & 2~\cite{Imshennik}& 11~\cite{KII} & 8~\cite{IMB} & 6~\cite{Baksan} \\
\hline
\end{tabular}
\end{center}
\end{table}

In this paper we discuss only the first stage of the presumable two-stage collapse and, accordingly, only the first neutrino signal, which occured at 2:52 UT. Moreover, we compare only LSD and KII signals, and do not use IMB and BST data. Even such restricted analysis leads to very rigid constraints on the first stage of the collapse, and additional data do not affect the conclusions significantly.


Time sequence and {\it measured} energies of LSD and KII events are listed in Table 2. Also the energy threshold $E_{th}$ and the mean number of background events with energy greater than $E_{th}$ in a ten-second interval are given for each detector. It should be reminded that due to the problems with clocks in KII the timing of KII event sequence could be shifted within one minute \cite{KII2}. According to \cite{DeRujula} a pair of KII events given in Table 2 is the largest cluster of events in KII at 2:52$\pm 1$min.

The imitation rate from the background for the LSD event cluster was fairly small - 0.7 per year \cite{LSD1},\cite{LSD2}. 
This justifies the attempts to find an explanation for the LSD neutrino signal. 

\begin{table}[t]
\caption{Energy thresholds $E_{th}$ and mean numbers of background events in a ten-second  
 interval $\bar n _{bg}$ according to the cited references; time sequence ($t$ is time in seconds) and energies $E$ 
  of  events at 2:52 for LSD \cite{LSD1}\cite{LSD2} and KII \cite{DeRujula}.}
\begin{center}
\begin{tabular}{|l|c|c|}
\hline
 & LSD & KII\\
\hline
$E_{th}$ (MeV) & 5 \cite{LSD1}\cite{LSD2}  & 7.5 \cite{KII2} \\
\hline
$\bar n _{bg}$ & 0.1 \cite{LSD1}\cite{LSD2} & 0.22 \cite{KII}\cite{KII2} \\
\hline
events 
& \begin{tabular}{cc}
         $t$ (s) & $E$ (MeV)\\[5pt]
         36.8 & 7\\
         40.7 & 8 \\
         41.0 & 11\\
         42.7 & 7 \\
         43.8 & 9
         \end{tabular}
&\begin{tabular}{cc}
         $t$ (s) & $E$ (MeV)\\[5pt]
         22 & 12\\
         30 & 7.5 \\
&\\
&\\
&
         \end{tabular}\\
\hline
\end{tabular}
\end{center}
\end{table}

\subsection{Reactions responsible for the neutrino registration in LSD and KII}
Neutrinos and antineutrinos can be registered through their interactions with nuclei and electrons.
Reactions which were responsible for neutrino and antineutrino detection in LSD and KII are listed in Table 3.
In this Table $l=e,\mu,\tau;~x=\mu,\tau,$ and superscript "*" denotes the exited states of the nuclei (which immediately decay to ground ones emitting nucleons and gammas). We have listed only those reactions which proved to be essential in the analysis of KII and LSD signals. Thus we omitted interactions of electron antineutrinos with nuclei and electrons because they could at best produce ten times less events than reaction (1) (see references in Table 5). Also we omitted all neutral current reactions with oxygen because KII, being a Cherenkov detector, was not sensitive to them.

\begin{table}[p]
\caption{Reactions responsible for the neutrino registration in LSD and KII. In the last two columns the 
detector efficiencies are listed and corresponding references are cited. 
They correspond to neutrino energies greater than 30 MeV for reactions (4), (7) and (10), 
and to energies greater than 15 MeV for other reactions. See text for further explanations and references.}
\begin{center}
\begin{tabular}{|l|rcl|c|c|}
\hline
 & \multicolumn{3}{|c|}{reaction} & $\eta$(LSD) & $\eta$(KII)\\
\hline
(1) & $\antinue + p$ & $\ar$ & $e^+ + n$ & 1 \cite{LSD1} & 0.9 \cite{KII2} \\ 
\hline
(2) & $\nue + \Fe $ & $\ar$ & $e + \Co^*$ & 0.53 \cite{Imshennik} & -- \\ 
\hline
(3) & $\nue + \C$ & $ \ar $ & $ e + \N^* $ & 1 \cite{LSD1} & -- \\ 
\hline
(4) & $\nue + \Ox$ & $\ar$ & $e + \F^*$ & -- & 0.9 \cite{KII2} \\ 
\hline
(5) & $ \nul + \Fe $ & $ \ar$ & $ \nul + \Fe^* $ & 0.53 \cite{Imshennik} & -- \\ 
\hline
(6) & $ \nul + \C$ & $ \ar $ & $ \nul + \C^* $ & 1 \cite{LSD1} & -- \\ 
\hline
(7) & $ \nul + e$ & $ \ar $ & $\nul + e$ & 1 \cite{LSD1} & $>$0.5 \cite{KII2} \\ 
\hline
(8) & $\antinux + \Fe$ & $ \ar $ & $ \antinux + \Fe^*$ & 0.53 \cite{Imshennik} & -- \\ 
\hline
(9) & $\antinux + \C$ & $ \ar $ & $ \antinux + \C^*$ & 1 \cite{LSD1} & -- \\ 
\hline
(10) & $ \antinux + e$ & $ \ar $ & $\antinux + e$ & 1 \cite{LSD1} & $>$0.5 \cite{KII2} \\ 
\hline
\end{tabular}
\end{center}
\end{table}

\begin{table}[p]
\caption{Electron antineutrino detection effective area $S_{\antinue}$ (in $10^{-10}$ cm$^2$)  for KII and LSD.}
\begin{center}
\begin{tabular}{|l||l|l|l|l|l|l|l|}
\hline
 $E_{\antinue}-\Delta m_{np}-m_e$ (MeV) & 6 & 7 & 8 & 9 & 10 & 11 & 12 \\
\hline
 $S_{\antinue}$(LSD) & 0.30 & 0.39 & 0.50 & 0.62 & 0.75 & 0.89 & 1.05\\
\hline
 $S_{\antinue}$(KII) & 0.6 & 1.8 & 3.6 & 6.0 & 9.1 & 11.6 & 14.4 \\
\hline
\end{tabular}
\end{center}
\end{table}

\begin{table}[p]
\caption{Cross-sections for reactions (1)-(6) (in $10^{-40}$cm$^2$) according   to the cited references.}
\begin{center}
\begin{tabular}{|l|l|l|l|l|l|l|}
\hline
 $E_{\nu}$(MeV) & (1) \cite{Strumia} & (2) \cite{Langanke} & (3) \cite{Fukugita}& (4) \cite{Haxton} & 
 (5) \cite{Langanke}& (6) \cite{Fukugita}\\
\hline
 15 & 0.162 & 0.064 & 0 & 0.001 & 0.021 & 0 \\
\hline
 20 & 0.290 & 0.293 & 0.003 & 0.002 & 0.069 & 0.003 \\
\hline
 25 & 0.447 & 0.733 & 0.019 & 0.008 & 0.151 & 0.012 \\
\hline
 30 & 0.630 & 1.40 & 0.050 & 0.02 & 0.285 & 0.027 \\
\hline
 35 & 0.835 & 2.36 & 0.095 & 0.05 & 0.489 & 0.048 \\
\hline
 40 & 1.058 & 3.71 & 0.151 & 0.12 & 0.786 & 0.073 \\
\hline
 45 & 1.298 & 5.55 & 0.218 & 0.24 & 1.19 & 0.101 \\
\hline
 50 & 1.551 & 7.98 & 0.292 & 0.42 & 1.72 & 0.131 \\
\hline
 55 & 1.816 & 11.0 & 0.370 & 0.67 & 2.39 & 0.162 \\
\hline
 60 & 2.091 & 14.8 & 0.452 & 1.0 & 3.20 & 0.195 \\
\hline
 65 & 2.374 & 19.2 & 0.530 & 1.46 & 4.15 & 0.224 \\
\hline
 70 & 2.664 & 24.2 & 0.608 & 2.20 & 5.25 & 0.254\\
\hline
\end{tabular}
\end{center}
\end{table}

It should be stressed that while in case of reaction (1) energy {\it measured} in the detector (i.e. energy of the outcoming positron) almost equals the incident neutrino energy, it is not the case for other reactions, especially for those which are accompanied by the nuclear excitations. In fact, for LSD it could be exclusively  nuclear excitation energy which was measured. This issue is discussed in detail in \cite{Imshennik},\cite{Gaponov}.

\newpage
Let us define an effective area $S_{\nul(\antinul)}$ for neutrino $\nul$ (or antineutrino $\antinul$) detection:
$$S_{\nul(\antinul)}\equiv \underset{i}\Sigma \eta_i N_i \sigma_i.$$
Here again $l=e,\mu,\tau,$ while $i$ numerates those reactions from the above list which are relevant for $\nul$ (or $\antinul$) detection, 
$\eta_i$ is the detector efficiency specific for the $i$-th reaction, $N_i$ is the number of corresponding target particles (electrons, protons or nuclei) in the detector and $\sigma_i$ is the corresponding cross section. Effective area depends on energy via cross sections and efficiencies. Evidently, mean number of registered neutrinos of a given type is proportional to the effective area of the detector. Thus effective area is a measure of the detector sensitivity.

First let us briefly discuss the possibility that LSD signal was due to the flux of electron antineutrinos, $\antinue,$
with energies in the range 
$6 ~{\rm MeV} < E_{\antinue}-\Delta m_{np}-m_e < 12 ~{\rm MeV}$, which corresponds to measured event energies.
 Here $\Delta m_{np}=1.3$ MeV is the the neutron-proton mass difference, and 
$m_e=0.5$ MeV -- the electron mass. Effective areas for $\antinue$ detection in LSD and KII through the reaction (1)  are given in Table 4. The efficiency for KII, which strongly depended on energy in this energy range, may be found in \cite{KII2}. Efficiency for LSD is considered to be 99\% according to \cite{LSD2}. The cross section for the reaction (1) is taken (here and in what follows) from \cite{Strumia} (eq. (25)). To interpret the data one should remember, that energy measured in LSD contained 1 MeV from $e^+e^-$ annihilation additionally to positron kinetic energy $E_{\antinue}-\Delta m_{np}-m_e $. 

Numerous studies (see, for example,  \cite{KII}, \cite{Dokuchaev}, \cite{Aglietta}, \cite{Dadykin}) indicated that if LSD signal were due to (7-12) MeV electron antineutrino flux, then\\
(1) expected average number of events in KII would be from 5\cite{Dadykin} to 28\cite{KII}, depending on the analysis, and \\
(2) an enormous energy $\gtrsim 10^{54}$erg, which is greater than binding gravitational energy of a neutron star, would be released during the collapse.\\
 Physical reasons, underlying difficulties in such an interpretation of the LSD data, are discussed in detail in \cite{Dokuchaev}\footnote{In \cite{Dokuchaev} a two-stage collapse scenario was proposed, which allowed to eliminate this difficulties, but implied an extremely heavy ($\gtrsim 10 M_\odot$) collapsing core.}. I what follows we do not consider this interpretation.

In 1987 only reaction (1) was regarded to allow supernova neutrino detection. Another possibility, on which we will focus our attention, was elaborated only in 2004 \cite{Imshennik}\cite{Gaponov}: the idea was that neutrinos (not antineutrinos) of sufficiently high (30-50 MeV) energy produced the signal in LSD through reactions on iron and carbon nuclei. 

Efficiencies for major reactions relevant for the detection of neutrinos and antineutrinos of such energies are listed in the last two columns of Table 3~\footnote{To estimate the efficiency of $\nu_l$ and $\antinux$ detection in KII due to the elastic scattering on electrons (reactions (7) and (10)) one should take into account the spectrum of scattered electrons.}. 
Cross sections for reactions (1) - (6) 
in the energy range (15-70) MeV are listed in Table 5. 
Effective areas of LSD and KII for neutrino detection are given in columns 2--5 of  Table 6. 
  In columns 6 and 7 one can see effective areas for the detection of high energy (with kinetic energy $>14$ MeV) electron from the $\nu_e e$ and $\antinux e$ elastic scattering  in KII. No such electrons were registered in KII which should be taken into account in the statistical analysis. The last column in Table 6 provides $\antinue$ detection area for KII, which is usefull to constrain $\antinue$ flux in the discussed energy range. At energies higher than 15 MeV $\antinue$ detection area of LSD is 16.5 times less than those of KII.


\begin{table}[t]
\caption{
Effective areas (in $10^{-10}$ cm$^2$) for neutrino (columns 2-5)
 and electron antineutrino (column 8)
  detection, $S_{\nul}$ and $S_{\antinue},$  for KII and LSD; effective areas for the detection of high energy (with kinetic energy $>14$ MeV) electron from the $\nu_e e$ and $\antinux e$ elastic scattering  in KII (columns 6 and 7). See text for further explanations.}
\begin{center}
\begin{tabular}{|l|l|l|l|l|l|l|l|}
 \hline
 $E_{\nu}$ (MeV) & $\nu _e,$ \text{ LSD} & $\nu _e,$ \text{ KII} & $\nu _y$\text{, LSD} & $\nu _y$\text{, KII}
& \text{HE }$\nu _e e,$\text{ KII} & \text{HE }$\nu _x e,$\text{ KII}&$\antinue$,\text{KII} \\
\hline 
 15 & \text{0.095} & \text{0.99} & \text{0.024} & \text{0.16} & \text{0.060} & \text{0.0085}
& \text{28} \\ \hline
 20 & \text{0.42} & \text{1.4} & \text{0.087} & \text{0.21} & \text{0.36} & \text{0.052}
& \text{51} \\  \hline
 25 & \text{1.1} & \text{2.1} & \text{0.21} & \text{0.26} & \text{0.67} & \text{0.097}
& \text{79} \\ \hline
 30 & \text{2.1} & \text{3.2} & \text{0.42} & \text{0.31} & \text{0.97} & \text{0.14}
& \text{114} \\ \hline
 35 & \text{3.7} & \text{5.6} & \text{0.72} & \text{0.36} & \text{1.3} & \text
{0.19}
& \text{155} \\ \hline
 40 & \text{5.8} & \text{10.4} & \text{1.1} & \text{0.41} & \text{1.59} & \text{0.24}
& \text{202} \\ \hline
 45 & \text{8.6} & \text{18.6} & \text{1.7} & \text{0.46} & \text{1.9}
& \text{0.29} & \text{256} \\ \hline
 50 & \text{12.3} & \text{30.8} & \text{2.4} & \text{0.51} & \text
{2.2} & \text{0.34}
& \text{316} \\ \hline
 55 & \text{16.8} & \text{47.6} & \text{3.2} & \text{0.57} & \text
{2.5} & \text{0.39}
& \text{382} \\ \hline
 60 & \text{22.3} & \text{69.7} & \text{4.3} & \text{0.62} & \text{2.8} & \text{0.44}
& \text{455} \\ \hline
 65 & \text{28.6} & \text{101} & \text{5.4} & \text{0.67} & \text{3.1} & \text{0.49}
& \text{534} \\ \hline
 70 & \text{35.7} & \text{150} & \text{6.7} & \text{0.72} & \text
{3.5} & \text{0.54}
& \text{619}\\
\hline
\end{tabular}
\end{center}
\end{table}

Unfortunately, we are not aware of any paper in which the cross-section of the antineutrino-iron reaction (8) is tabulated. That is why
we can not calculate non-electron antineutrino LSD effective area. However, cross-sections of interactions of 
{\it antineutrinos} with heavy nuclei are in general not greater than those of {\it neutrinos} (see, for example, \cite{Langanke}). Therefore we may assert that non-electron antineutrino LSD effective area was not greater than electron antineutrino LSD effective area:
\be\label{areas}
S_{\antinux}\leq S_{\nux}.
\ee

It is easily seen from the Table 6 that\\
(1) KII was roughly an order of magnitude more sensitive to $\antinue$ than to $\nue$ and $\nux;$\\
(2) electron neutrino ($\nu_e$) sensitivities of LSD and KII were comparable in the energy range $E\simeq (25-45)$  Mev;\\
(3) non-electron neutrino ($\nu_x$) sensitivity of LSD was greater then those of KII for energy $\gtrsim 30$ MeV.\\
At this stage one may think that no contradiction between LSD and KII event numbers would occur if supernova neutrino flux was composed of electron and non-electron neutrinos of appropriate energies. The question is weather such flux content is possible in principle. To answer this question one should consider neutrino emission by the collapsing supernova core and neutrino flavour transformation in the matter of the star due to MSW-effect.  

\section{Supernova neutrino emission and flavour transformations}

There is a number of reviews on the supernova neutrino emission and flavour transformations (see, for example, 
\cite{Takahashi review}, \cite{Burrows review}). In this section we
 utilise the results described there in detail.

Neutrinos and antineutrinos of all three flavours can be created during the collapse of the iron core in the center of the star. All reactions in which they are created conserve lepton flavour. In particular, muon neutrinos and antineutrinos are created only in pairs, the same is 
valid for tau-neutrinos and antineutrinos. Moreover, muon neutrinos are produced in the same reactions as tau-neutrinos. Therefore in {\it any} collapse model neutrino fluxes satisfy the following conditions:

\be\label{F0 restriction 1}
F^0_x \equiv
F^0_{\nu_{\mu}}=F^0_{\nu_{\tau}},~~~~~F^0_{\bar x}\equiv F^0_{{\bar{\nu}_{\tau}}}=F^0_{{\bar{\nu}_{\mu}}},
~~~~~ F^0_{x}=F^0_{\bar x},
\ee 

\be\label{F0 restriction 2}
0\leq F^0_{\nu_e}-F^0_{\bar{\nu}_e}\leq \frac{N^{\rm Core}_{\rm e}}{4\pi L^2}
\ee 
Here $F_{\nul,\antinul}^0$ is a time- and energy-integrated flux of an (anti-)neutrinos, subscript specifying the type of (anti-)neutrinos.
 $F^0_x$ is a standard notation for non-electron neutrino original flux. Superscript "0" denotes that it is an original flux, i.e. such a flux which would reach the earth if there were no flavour transformations in the matter of the star. $L=52$kpc is the distance between the supernova and the earth, and $N^{\rm Core}_{\rm e}$ is the number of electrons in the iron core. 
\be
N^{\rm Core}_{\rm e}=26 \frac{M_{\rm Core}}{m_{\Fe}},
\ee
where $m_{\Fe}$ is the iron nucleus mass, and 26 is the number of electrons in the atom of iron.

 The supernova collapse begins when the iron core reaches the Chandrasekhar mass limit, $M_{\rm Ch}.$  Chandrasekhar mass varies from $1.4M_\odot$ to $2.2M_\odot,~M_\odot$ being the mass of the Sun (see, for example, review \cite{Imshennik and Nadyozhin} or
 paper \cite{about M_Ch}).
 Last inequality in (\ref{F0 restriction 2}) reads
\be\label{max flux}
0 \leq F^0_{\nu_e}-F^0_{\bar{\nu}_e} \leq  0.34\frac{M_{\rm Core}}{2M_\odot } \cdot 10^{10}{\rm cm}^{-2}
\ee

Due to the MSW-effect \cite{W}\cite{MS} neutrinos may change their flavour while passing through the envelope of the star. In general neutrino 
and antineutrino fluxes at the earth $F_{\nul,\antinul}$ are linear combinations of original fluxes \cite{Smirnov}\cite{Takahashi}:
\be F_{\nu_e}=pF^0_{\nu_e} + (1-p)F^0_x, \label{01}\ee
\begin{equation}\label{02}
F_{\bar{\nu}_e}=\bar{p}F^0_{\bar{\nu}_e}+(1-\bar{p})F^0_x,
\end{equation}
\be\label{03}
F_{\nu_\mu}+F_{\nu_\tau}=(1-p)F^0_{\nu_e} + (1+p)F^0_x,
\ee
\be\label{04}
F_{\bar\nu_{\mu}}+F_{\bar\nu_{\tau}}=(1-\bar{p})F^0_{\antinue}+(1+\bar{p})F^0_x,
\ee
where 
conversion probabilities
$p,~\bar p$  depend on the unknown neutrino mass hierarchy and neutrino mixing angle $\theta_{13}.$ One may distinguish three extreme (with respect to $\theta_{13}$) cases: \\
{\bf LN} ( {\bf
L}arge angle, {\bf N}ormal hierarchy),\\
{\bf LI} ({\bf L}arge
angle, {\bf I}nverted hierarchy),\\
{\bf SA} ({\bf S}mall angles, {\bf A}ny hierarchy).\\
 Here
large angles stand for $\theta_{13}\gtrsim 3\cdot 10^{-2}$, while
small angles stand for $\theta_{13}\lesssim 3\cdot
10^{-3}$. In addition a case of intermediate $\theta_{13}$ should be considered. Note that current experimental limit is $\theta_{13}<0.17, ~~\sin^2\theta_{13}<0.03$ ~\cite{CHOOZ}.
Coefficients for all this cases are given \newline  in {Table 7} \cite{Smirnov}\cite{Takahashi}.

\begin{table}[t]
\caption{Conversion probabilities $p$ and $\bar p$ for different neutrino mass hierarchies 
and values of $\theta_{13}.$}
\begin{center}
\begin{tabular}{|l||c|c|c|}
\hline
 & $\theta_{13}\lesssim 3\cdot
10^{-3}$ & $\theta_{13}\gtrsim 3\cdot 10^{-2}$ & $3\cdot
10^{-3}\lesssim \theta_{13}\lesssim 3\cdot 10^{-2}$ \\
\hline
Normal hierarchy &  & $p= 0$ 
& $0 \leq p\leq \sin^2 \theta_{12}$\\
  & $p= \sin^2 \theta_{12}$ & $\bar p= \cos^2 \theta_{12}$
& $\bar p= \cos^2 \theta_{12}$ \\
\cline{1-1}\cline{3-4}
Inverted hierarchy & $\bar p= \cos^2 \theta_{12}$ & $p= \sin^2 \theta_{12}$ 
& $p= \sin^2 \theta_{12}$ \\
  & & $\bar p= 0$ 
& $0 \leq \bar p\leq \cos^2 \theta_{12}$ \\
\hline
\end{tabular}
\end{center}
\end{table}

According to PDG \cite{PDG} $\sin^2 \theta_{12}=0.28.$

Two consequences follow immediately from eq.(\ref{02}) and Table 7.\\

(1) Upper bound on  $F^0_x:$

\be\label{antinueF0 constraint}
F^0_x \leq \frac{1}{1-\bar p}F_{\bar{\nu}_e}\leq 3.5 F_{\bar{\nu}_e}
\ee

Thus the original flux $F^0_x$ was constrained by the flux at the earth $F_{\bar{\nu}_e}$, which by itself was severely constrained in the considered energy range by KII non-observation of electron antineutrinos with $E>14$MeV.\\

(2) Upper bound on  $F^0_{\antinue}$ in  \LN and \SA cases:

\be\label{antinueF0 constraint}
F^0_{\antinue} \leq \frac{1}{\bar p}F_{\bar{\nu}_e}=
1.4 F_{\bar{\nu}_e}, ~~~~ \text{\LN~ and \SA}
\ee

Thus if either \LN  or \SA  case is realised in Nature then the original flux $F^0_{\antinue}$ was constrained for the same reason as $F^0_x.$ 
In the \LI case $F_{\antinue}$ is independent from $F^0_{\antinue}.$

Numerical values for both upper bounds are given in the next section (eqs. (\ref{F0 restr 1}) and (\ref{F0 restr 2})).\\

Values for $p, \bar p$ given in Table 7 are valid if the density profile of the envelope of the progenitor star is static during neutrino emission time interval ($\sim 10$ seconds).  This assumption may be violated by a shock wave, which moves from the core through the envelope. If it reaches the region of neutrino conversion (resonance region), it may change values of $p$ and $\bar p$ \cite{Schirato and Fuller}.
  It is a shock wave which blows up the star envelope, therefore it is an inevitable feature of the supernova explosion. However, it is not clear whether the {\it first} stage of the two-stage explosion should be accompanied by the shock wave powerful enough to reach the resonance region.

 Simple physical considerations and detailed numerical studies \cite{Schirato and Fuller},\cite{Takahashi shock wave},\cite{Smirnov shock wave},\cite{Tomas} show that \\
(1) shock wave reaches the resonance region and in 2-5 seconds after the onset of the collapse;\\
(2) in the \SA case the shock wave effect is negligible;\\
(3) in the \LN case shock wave switches $p$ from 0 to some non-zero value in the interval 
$\lbrack 0,\sin^2 \theta_{12} \rbrack,$ but does not affect $\bar p;$\\
(4) in the \LI case shock wave switches $\bar p$ from 0 to some non-zero value in the interval 
$\lbrack 0,\cos^2 \theta_{12} \rbrack,$ but does not affect $p.$\\

SN1987A emerged in the Southern hemisphere, and the neutrinos had to pass through the earth in order to reach LSD and KII. The earth matter effect for supernova neutrinos is described in detail for example in \cite{Smirnov shock wave}. Not going into details we note that in general it could change $p$ and $\bar p$ by 5\%- 30\%, but in the \LN case it could not affect $p,$ and in the \LI case -- $\bar p.$ 
As a result the earth matter effect proved not to influence significantly the results of the statistical analysis described in the next section.

\section{Statistical analysis and results}

Energy distribution of supernova neutrinos is thought to be some modification of Fermi-Dirac distribution (see, for example, \cite{Imshennik and Nadyozhin}). Unfortunately, it is absolutely impossible to reconstruct the shape of the distribution from the LSD neutrino signal because as it was stressed above the {\it measured} event energy carried almost no information about the incident neutrino energy, to say nothing about small number of events. Therefore in our statistical analysis we consider a set of "monoenergetic" hypothesises. To be more precise, we numerically investigate the probability $P(E,F^0_{\nu_e},F^0_{\antinue},F^0_{\nu_x,})$ that monoenergetic (with energy $E$) fluxes $F^0_{\nu_e},$ $F^0_{\antinue}$ and $F^0_x$ after the flavour transformation according to eqs.(\ref{01})-(\ref{04}) produced \\
(1) not less than 5 events in LSD, \\
(2) not greater than 2 events in KII, no one from which being a $\nu e$ elastic scattering with $E_e>14$ MeV. \\
 Poisson statistics is applied. Energy runs from 15 to 70 MeV with the step 5 MeV.  Three combinations of neutrino mass hierarchy and mixing angle $\theta_{13}$, described in the previous section, are implemented.

First we maximize $P(E,F^0_{\nu_e},F^0_{\antinue},F^0_x)$ with respect to $F^0_{\nu_e},$ fixing each of the other two original fluxes to be one of the values from the set $\{(0.5,~1,~1.5,~2) \times  10^8$cm$^{-2}\}.$
We find that\\
1. the maximal value of probability $P$ does not exceed $1\%$ for values of $F^0_{x}$ greater than $2\cdot10^8$cm$^{-2}$ 
 for all combinations of neutrino parameters (including extreme cases \LN, \LI, \SA):
\be \label{F0 restr 1}
F^0_x<2 \cdot 10^8{\rm cm}^{-2}, ~~~ 99\% {\rm CL};
\ee
2. the maximal value of probability $P$ is less  than $1\%$ for values of $F^0_{\antinue}$ greater than $10^8$cm$^{-2}$ 
for the \LN and \SA cases:
\be \label{F0 restr 2}
F^0_{\antinue}<10^8{\rm cm}^{-2}, ~~~ 99\% {\rm CL};
\ee
Combining this with inequality (\ref{max  flux}) one obtains an upper bound on $F^0_{\nue}:$
\be\label{max flux 2}
F^0_{\nu_e} \leq  0.34\frac{M_{\rm Core}}{2M_\odot } \cdot 10^{10}{\rm cm}^{-2}, \LN \text{or} ~\SA.
\ee


Simple estimates of the mean number of events in the detectors, as well as final results, confirm that the value $10^8$cm$^{-2}$ is negligible compared to the original neutrino flux sufficient to explain LSD signal, which is of order of $10^{10}$cm$^{-2}$. Thus, in the subsequent analysis we may safely put $F^0_x=0$ and, in the \LN and \SA cases, 
$F^0_{\antinue}=0.$ In order to consider the \LI case on equal footing with the \LN and \SA cases, we note that according to (\ref{areas})
\be
 P(E,F^0_{\nu_e},F^0_{\antinue},0)<P(E,F^0_{\nu_e}+F^0_{\antinue},0,0).
\ee
  This allows to consider $P(E,F^0_{\nu_e}+F^0_{\antinue},0,0)$ as an upper bound on  $P(E,F^0_{\nu_e},F^0_{\antinue},0)$ in the \LI case.

After this simplifications we maximize probability $P(E,F^0,0,0)$ with respect to effective original flux $F^0,$ which stands for $F^0_{\nu_e}$ in the \LN and \SA cases and for $F^0_{\nu_e}+F^0_{\antinue}$ in the \LI case. When the maximal probability $P_{\rm max}$ is greater than 2\% we find lower and upper bounds for $F^0,~ F_{\rm upper}^0$ 
and $F_{\rm lower}^0$ correspondingly, solving equation $P(E,F^0,0,0)=0.02.$ 
The result for the \LI and \SA cases is presented in Table 8, 
and for the \LN case -- in Table 9. Only those energies, for which $P_{\rm max}>2\%,$ are included in the tables.

\begin{table}[t]
\caption{Maximal probability, lower and upper bounds and central value for the effective \tabularnewline original flux  in the \LI and \SA cases. Fluxes are measured in $10^{10}$ cm$^2.$
 See text for further explanations.}
\begin{center}
\begin{tabular}{|l|c|c|c|c|}
\hline
$E_{\nu}(MeV)$ & $P_{\rm max}$ & $F^0_{\rm lower}$ & $F^0_{\rm central}$ & $F^0_{\rm upper}$ \\
\hline
 30 & 0.030 & 2.2 & 3.4   & 5.0\\
\hline
 35 & 0.037 & 1.2  & 2.1  & 3.3 \\
\hline
 40 & 0.032 & 0.77  & 1.2  & 1.8 \\
\hline
 45 & 0.024 & 0.57 & 0.75   & 0.97 \\
\hline
\end{tabular}
\end{center}
\end{table}

\begin{table}[t]
\caption{
Maximal probability, lower and upper bounds and central value for the effective 
 original flux  in the \LN case. Fluxes are measured in $10^{10}$ cm$^2.$ The lower block
 of the table correspond to the flux allowed by inequality (\ref{max flux 2}) 
with $M_{\rm Core}=2 M_{\odot}$. 
See text for further explanations.
}
\begin{center}
\begin{tabular}{|l|c|c|c|c|}
\hline
$E_{\nu}(MeV)$ & $P_{\rm max}$ & $F^0_{\rm lower}$ & $F^0_{\rm central}$ & $F^0_{\rm upper}$ \\
\hline
\hline
 25 & 0.029 & 9.6 &   14.7   & 21.2 \\
\hline
 30 & 0.075 & 4.1 &   9.7  & 19.5 \\
\hline
 35 & 0.13 & 2.3 &   6.7   & 16.3 \\
\hline
 40 & 0.19 & 1.4 &   4.8   & 13.5 \\
\hline
 45 & 0.25 & 0.91 &   3.5   & 10.9 \\
\hline
 50 & 0.31 & 0.64 &   2.7  & 8.7 \\
\hline
 55 & 0.36 & 0.46 &   2.1   & 6.9 \\
\hline
 60 & 0.40 & 0.35 &   1.64   & 5.6 \\
\hline
\hline
 65 & 0.44 & 0.27 &   1.34  & 4.6 \\
\hline
 70 & 0.48 & 0.22 &   1.11   & 3.8\\
\hline
\end{tabular}
\end{center}
\end{table}

\begin{table}[p]
\caption{Maximal probability, lower and upper bounds and central value for the  effective 
 original flux  in the \LI and \SA cases, in case of 1.3 times increased LSD and 1.4 times  decreased KII effective areas. Fluxes are measured in $10^{10}$ cm$^2.$ The lower block
 of the table correspond to the flux allowed by inequality (\ref{max flux 2}) 
with $M_{\rm Core}=2 M_{\odot}$. 
See text for further explanations.}
\begin{center}
\begin{tabular}{|l|c|c|c|c|c|c|}
\hline
$E_{\nu}(MeV)$ & $P_{\rm max}^0$ & $F_{\rm lower~2\%}^0$ & $F_{\rm lower ~5\%}^0$ & $F_{\rm central}^0$ &
 $F_{\rm upper~5\%}^0$
& $F_{\rm upper~2\%}^0$ \\
\hline
\hline
 25 & 0.031 & 3.4 & - & 5.3 & - & 7.8 \\
\hline
 30 & 0.066 & 1.47 & 2.3 & 3.3 & 4.5 & 6.2 \\
\hline
 35 & 0.089 & 0.82 & 1.20 & 2.0 & 3.2 & 4.2 \\
\hline
 40 & 0.092 & 0.51 & 0.74 & 1.27 & 2.0 & 2.6\\
\hline
\hline
 45 & 0.082 & 0.34 & 0.50 & 0.81 & 1.2 & 1.6 \\
\hline
 50 & 0.070 & 0.24 & 0.37 & 0.54 & 0.75 & 1.01 \\
\hline
 55 & 0.058 & 0.18 & 0.29 & 0.37 & 0.47 & 0.67 \\
\hline
 60 & 0.049 & 0.14 & - & 0.27 & -  & 0.46 \\
\hline
 65 & 0.039 & 0.11 & - & 0.20 & -& 0.31 \\
\hline
 70 & 0.026 & 0.10 & - & 0.14 & - & 0.19\\
\hline
\end{tabular}
\end{center}
\end{table}

\begin{table}[p]
\caption{Maximal probability, lower and upper bounds and central value for the  effective 
 original flux  in the \LN case, in case of 1.3 times increased LSD and 1.4 times  decreased KII effective areas. Fluxes are measured in $10^{10}$ cm$^2.$ The lower block
 of \tabularnewline   the table correspond to the flux allowed by inequality (\ref{max flux 2}) 
with $M_{\rm Core}=2 M_{\odot}$.
See text for further explanations.}
\begin{center}
\begin{tabular}{|l|c|c|c|c|c|c|}
\hline
$E_{\nu}(MeV)$ & $P_{\rm max}$ & $F_{\rm lower ~2\%}^0$ & $F_{\rm lower ~5\%}^0$ & $F_{\rm central}^0$ & 
$F_{\rm upper ~5\%}^0$
& $F_{\rm upper ~2\%}^0$ \\
\hline
\hline
 25 & 0.055 & 6.4 & 10.9 & 13.2 & 15.9 & 23.8 \\
\hline
 30 & 0.12 & 3.0 & 4.2 & 8.5 & 15.5 & 19.8\\
\hline
 35 & 0.19  & 1.7 & 2.3 & 5.8 & 12.9 & 16.0 \\
\hline
 40 & 0.26  & 1.0 & 1.4  & 4.0  & 10.5  & 12.6 \\
\hline
 45 & 0.33 & 0.69 & 0.92  & 2.9  & 8.3  & 9.6  \\
\hline
 50 & 0.38 & 0.48 & 0.64 & 2.2 & 6.5 & 7.4 \\
\hline
 55 & 0.44 & 0.35 & 0.47 & 1.70 & 5.1 & 5.8 \\
\hline\hline
 60 & 0.48  & 0.27 & 0.35 & 1.35 & 4.1 & 4.6\\
\hline
 65 & 0.52 & 0.21 & 0.28 & 1.10 & 3.3 & 3.7 \\
\hline
 70 & 0.55 & 0.17 & 0.22 & 0.91 & 2.8 & 3.1\\
\hline
\end{tabular}
\end{center}
\end{table}

The following conclusions may be derived from this tables\\

\SA case. Those values of $F^0_{\antinue},$ which agree with the data (see Table 8), do not fill inequality (\ref{max flux 2}). The shock wave effect, as noted in the previous section, is negligible.\\

\LI case. Electron neutrinos and antineutrinos in the energy range (30-45) MeV with the total flux of order of 
$10^{10}$ cm$^2$ fit the data. Inequality (\ref{max flux 2}) does not hold for this case, the zero value of $\bar p=0$ being the reason for this. However, a powerful  shock wave could switch $\bar p$ to non-zero value, thus making  inequality (\ref{max flux 2}) obligatory for this case also. Therefore shock wave could spoil the agreement of the theory with the data. \\

\LN case. Taking into account inequality (\ref{max flux 2}) with $M_{\rm Core}=2 M_{\odot}$, one may conclude that electron neutrinos with energies greater than 60-65 MeV with the total flux of order of 
$10^{10}$ cm$^2$ fit the data. A powerful shock wave could switch $p$ from 0 to some nonzero value, thus effectively switching the \LN case to some modification of the \SA case. This could spoil the agreement of the theory with the data.\\

Maximal value of probability in Table 8  does not exceed 4\%. Table 9 and inequality (\ref{max flux 2}) imply unusually high neutrino energies. Evidently, agreement of the theory with the data is far from being perfect. However, 
various possible sources of inaccuracy in the analysis may be suspected, for example the following.\\
(1) Cross section of neutrino-nucleus interactions used above are obtained mainly as a result of calculations,
 not measurements, and may appear to be inaccurate. Thus LSD effective area may appear to be greater, and KII effective area
-- smaller than in Table 6.\\
(2) The value of the efficiency of LSD neutrino event reconstruction which we utilised according to \cite{Imshennik},
 53\%, is claimed to be an estimate. If it is greater than  53\%, then the LSD effective area increases.\\

To show that our results are stable under the reasonable variation of the input, we repeated calculations with the increased 
LSD effective area (by factor 1.3) and decreased KII effective area (by factor 1.4). We believe that such a procedure accounts for possible input variations in an optimistic (with respect to the concordance of the LSD signal with the double-burst hypothesis) way.
The results are presented in Tables 10 and 11.  In this tables two pairs of upper and lower bounds are presented, one of which 
corresponding to $P=2\%,$ and another -- to $P=5\%.$ It is clear from this tables that most of our conclusions remain unchanged. 
Only the status of the \SA case changes from "excluded" to "disfavoured". Note that in all cases but one, the \LI case without shock wave, large mass of the iron core $M^{\rm Core}$ is favoured.


\section{Conclusions}

The following conclusions concerning the first stage of the two-stage SN1987A explosion models are obtained.\\
\newline
(1) In the case of small mixing angle $\theta_{13}$, $\theta_{13}<0.003,$ such models are disfavoured by the data, independently of the neutrino mass hierarchy. \\
\newline
(2)  In any model non-electron neutrino and antineutrino production had to be severely suppressed during the first stage of the collapse,
independently of the neutrino mass hierarchy and mixing angle $\theta_{13}:$
\be
F^0_x \lesssim 10^8{\rm cm}^{-2}. 
\ee
 This means that at the first stage of the collapse there was no thermal equilibrium, even rough. \\
\newline
(3) In the case of normal mass hierarchy and large mixing angle $\theta_{13}$, $\theta_{13}>0.03,$ in order to explain the data one should imply emission of very energetic  ($E\gtrsim 60$ MeV) electron neutrinos, $\nue,$  at the first stage of the explosion; at the same time the suppression of $\antinue$ production should be assumed:
\be
F^0_{\nue} \simeq (0.3-0.5) \cdot 10^{10}{\rm cm}^{-2},~~~~~~~~~~~~ F^0_{\antinue}\lesssim 10^8{\rm cm}^{-2}.
\ee
 In addition, large values of the collapsing core mass, $M_{\rm Core}\gtrsim 2 M_{\odot},$ are necessary. A powerful shock wave could further complicate the agreement of the data with the theory.
\\
\newline
(4) In the case of inverted mass hierarchy and large mixing angle $\theta_{13}$, $\theta_{13}>0.03,$ the data can be explained by the moderate energy (30 MeV$\lesssim E \lesssim 45$ MeV) electron neutrino and antineutrino emission at the first stage of the explosion with fluxes of order of  $10^{10}{\rm cm}^{-2}.$
 A powerful shock wave could worsen the agreement of the data with the theory. \\

The conclusions of the current work incorporate those obtained in the previous investigation \cite{Lychkovskiy} of the author. In particular, following \cite{Imshennik} and \cite{Gaponov}, in \cite{Lychkovskiy} the energy of the neutrinos
responsible for the LSD signal was supposed to be in the 30-50 MeV range; under this assumption it was concluded that the case of normal mass hierarchy and large mixing angle $\theta_{13}$ is not compatible with the data. This accords with our current conclusion (3).
\newline
\newline
{\bf\Large Acknowledgements}\\
 The author wishes to thank O.G. Ryazhskaya, V.S. Imshennik, V.I.Dokuchaev and L.B.
 Okun for valuable discussions and attention to this work. The
 work was financially supported by Dynasty Foundation scholarship, by RF
 President grant NSh-5603.2006.2 and by RFBR grants 07-02-00830-a and 05-02-17203-a.



\begin{thebibliography}{99}

\bibitem{LSD1}
  V.~L.~Dadykin {\it et al.},
  JETP Lett.\  {\bf 45}, 593 (1987)
  [Pisma Zh.\ Eksp.\ Teor.\ Fiz.\  {\bf 45}, 464 (1987)]

\bibitem{LSD2}
  M.~Aglietta {\it et al.},
  Europhys.\ Lett.\  {\bf 3}, 1315 (1987).



\bibitem{KII}
  K.~Hirata {\it et al.}  [Kamiokande-II Collaboration],
  Phys.\ Rev.\ Lett.\  {\bf 58}, 1490 (1987).


\bibitem{IMB}
  R.~M.~Bionta {\it et al.},
  Phys.\ Rev.\ Lett.\  {\bf 58}, 1494 (1987).



\bibitem{Baksan}
  E.~N.~Alekseev, L.~N.~Alekseeva, V.~I.~Volchenko and I.~V.~Krivosheina,
  JETP Lett.\  {\bf 45}, 589 (1987)
  [Pisma Zh.\ Eksp.\ Teor.\ Fiz.\  {\bf 45}, 461 (1987)]


\bibitem{Hillebrandt}
W. Hillebrandt, P. Hoflich, P. Kafka, E. Muller, H.U. Schmidt, J.W. Truran, 
Astronomy and Astrophysics, {\bf 180}, L20 (1987)

\bibitem{DeRujula}
  A.~De Rujula,
  Phys.\ Lett.\  B {\bf 193}, 514 (1987).



\bibitem{Stella}
		L. Stella, A. Treves, 	Astronomy and Astrophysics {\bf 185}, L5 (1987)

\bibitem{Dokuchaev}
V.S. Berezinsky, C. Castagnoli, V.I. Dokuchaev, P. Galeotti, Il Nuovo Cimento, {\bf C11}, 287(1988)

\bibitem{Imshennik}
  V.~S.~Imshennik and O.~G.~Ryazhskaya,
  Astron.\ Lett.\  {\bf 30}, 14 (2004)
  [arXiv:astro-ph/0401613]

\bibitem{W}
  L.~Wolfenstein,
  Phys.\ Rev.\ D {\bf 17}, 2369 (1978).

\bibitem{MS}
  S.~P.~Mikheev and A.~Y.~Smirnov,
  Sov.\ J.\ Nucl.\ Phys.\  {\bf 42}, 913 (1985)
  [Yad.\ Fiz.\  {\bf 42}, 1441 (1985)].

\bibitem{Smirnov and Lunardini}
  C.~Lunardini and A.~Y.~Smirnov,
  Astropart.\ Phys.\  {\bf 21}, 703 (2004)
  [arXiv:hep-ph/0402128].

\bibitem{Lychkovskiy}
  O.~Lychkovskiy,
  arXiv:hep-ph/0604113.

\bibitem{Schirato and Fuller}
  R.~C.~Schirato, G.~M.~Fuller, 
  arXiv:astro-ph/0205390.

\bibitem{KII2}
  K.~S.~Hirata {\it et al.},
  Phys.\ Rev.\  D {\bf 38}, 448 (1988).

\bibitem{Gaponov}
  Y.~V.~Gaponov, O.~G.~Ryazhskaya and S.~V.~Semenov,
  Phys.\ Atom.\ Nucl.\  {\bf 67}, 1969 (2004)

\bibitem{Strumia}
  A.~Strumia and F.~Vissani,
  Phys.\ Lett.\  B {\bf 564}, 42 (2003)
  [arXiv:astro-ph/0302055].

\bibitem{Aglietta}
  M.~Aglietta {\it et al.},
  Europhys.\ Lett.\  {\bf 3}, 1321 (1987).


\bibitem{Dadykin}
  V.~L.~Dadykin, G.~T.~Zatsepin and O.~G.~Ryazhskaya,
  Sov.\ Phys.\ Usp.\  {\bf 32}, 459 (1989)
  [Usp.\ Fiz.\ Nauk {\bf 158}, 139 (1989)].






\bibitem{Langanke}
  E.~Kolbe and K.~Langanke,
  Phys.\ Rev.\  C {\bf 63}, 025802 (2001)
  [arXiv:nucl-th/0003060].

\bibitem{Fukugita}
  M.~Fukugita, Y.~Kohyama and K.~Kubodera,
  Phys.\ Lett.\  B {\bf 212}, 139 (1988).


\bibitem{Haxton}
  W.~C.~Haxton,
  Phys.\ Rev.\  D {\bf 36}, 2283 (1987).


\bibitem{Takahashi review}
  K.~Kotake, K.~Sato and K.~Takahashi,
  Rept.\ Prog.\ Phys.\  {\bf 69}, 971 (2006)
  [arXiv:astro-ph/0509456].

\bibitem{Burrows review}
  A.~Burrows,
  Ann.\ Rev.\ Nucl.\ Part.\ Sci.\  {\bf 40}, 181 (1990).



\bibitem{Imshennik and Nadyozhin}
V. S. Imshennik and D. K. Nadyozhin, Usp. Fiz. Nauk
{\bf 156}, 576 (1988)


\bibitem{about M_Ch}
  K.~Langanke and G.~Martinez-Pinedo,
  Rev.\ Mod.\ Phys.\  {\bf 75}, 819 (2003)
  [arXiv:nucl-th/0203071].





\bibitem{Smirnov}
  A.~S.~Dighe and A.~Y.~Smirnov,
  Phys.\ Rev.\ D {\bf 62}, 033007 (2000)
  [arXiv:hep-ph/9907423].

\bibitem{Takahashi}
 K.~Takahashi and K.~Sato,
 Prog.\ Theor.\ Phys.\  {\bf 109}, 919 (2003)
 [arXiv:hep-ph/0205070].




\bibitem{CHOOZ}
  M.~Apollonio {\it et al.},
  Eur.\ Phys.\ J.\ C {\bf 27}, 331 (2003)
  [arXiv:hep-ex/0301017].



\bibitem{PDG}
  W.~M.~Yao {\it et al.}  [Particle Data Group],
  ``Review of particle physics,''
  J.\ Phys.\ G {\bf 33}, 1 (2006)










\bibitem{Takahashi shock wave}
  K.~Takahashi, K.~Sato, H.~E.~Dalhed and J.~R.~Wilson,
  Astropart.\ Phys.\  {\bf 20}, 189 (2003)
  [arXiv:astro-ph/0212195].

\bibitem{Smirnov shock wave}
  C.~Lunardini and A.~Y.~Smirnov,
  JCAP {\bf 0306}, 009 (2003)
  [arXiv:hep-ph/0302033].

\bibitem{Tomas}
  R.~Tomas, M.~Kachelriess, G.~Raffelt, A.~Dighe, H.~T.~Janka and L.~Scheck,
  JCAP {\bf 0409}, 015 (2004)
  [arXiv:astro-ph/0407132].



\end{thebibliography}
\end{document}